# Gender differences in the equity-efficiency trade-off

Valerio Capraro

Middlesex University London
V.Capraro@mdx.ac.uk

**Abstract**

Gender differences in human behaviour have attracted generations of social scientists, who have explored whether males and females act differently in domains involving competition, risk taking, cooperation, altruism, honesty, as well as many others. Yet, little is known about gender differences in the equity-efficiency trade-off. It has been suggested that females are more equitable than males, but the empirical evidence is weak and inconclusive. This gap is particularly important, because people in power of redistributing resources often face a conflict between equity and efficiency, to the point that this trade-off has been named as "the central problem in distributive justice". The recently introduced Trade-Off Game (TOG) – in which a decision-maker has to unilaterally choose between being equitable or being efficient – offers a unique opportunity to fill this gap. To this end, I analyse gender differences on a large dataset including all N=5,056 TOG decisions collected by my research group since we introduced this game. The results show that females prefer equity over efficiency to a greater extent than males do. These findings suggest that males and females have different preferences for resource distribution, and point to new avenues for future research.

**Introduction**

After the 2016 Central Italy earthquake, which destroyed dozens of small mountain villages, killed hundreds of people, and left tens of thousands of other people homeless, the Italian Government found itself in the middle of a fundamental decisional conflict. MPs had to decide in which areas to build the temporary houses to host the survivors who had lost their house as a consequence of the seism. These temporary accommodations were meant to host the survivors for a relatively long time, estimated to about 10 years, while waiting for the reconstruction of their houses. For this reason, the survivors had a strong preference for having these temporary houses built as near as possible to where they used to live before the quake. The conflict emerged because this "equitable" solution, which would have satisfied all the survivors, was impracticable from the Government point of view: eliminating the rubbles in a reasonable time frame, reaching nearly inaccessible mountain villages with the trucks, and build, in each of these villages, a relatively small number of houses (some of these villages had only 20 inhabitants), would have exponentially inflated the cost and the time needed for the intervention. From the point of view of the Government, the most "efficient" solution was to select one single area and build all the temporary houses in this area. However, this solution was perceived to be highly unequal from the survivors: it would have satisfied some of them (those who happened to live near the selected area), and it would have dissatisfied others (those who happened to live far).

This is only one example of the tension between equity and efficiency. A more classical case is taxation: according to Okun's "leaky bucket" argument, taxation is inefficient, as administering the tax has a cost for the institution implementing the tax (Okun, 2015). The problem is, in fact, much deeper and relies in the often-unavoidable discrepancy between the natural and the equitable distributions of resources: the natural distribution of resources is often unequal, and creating equity is often costly. This generates a fundamental conflict between equity and efficiency, which has been named as "the central problem of distributive justice" (Hsu, Anen & Quartz, 2008) or "the big tradeoff" (Okun, 2015). People in power of resource distribution often face this conflict. For this reason, understanding what individual factors affect this trade-off is a problem of primary importance across social sciences. In this paper, I focus on one particular but important factor: the gender of the decision maker.

Gender differences in human behaviour have attracted generations of social scientists, who have used economic games to explore whether males and females act differently in a number of domains, including competition (Gneezy, Niederle & Rustichini, 2003; Niederle & Vesterlund, 2007; Gneezy, Leonard & List, 2009), risk taking (Powell & Ansic, 1997; Byrnes, Miller & Schafer, 1999; Charness & Gneezy, 2012), cooperation (Rand, 2017), altruism (Engel, 2011; Rand et al., 2016; Brañas-Garza et al., 2018), honesty (Capraro, 2018; Gerlach, Teodorescu & Hertwig, 2019; Abeler, Nosenzo & Raymond, in press), as well as many others (Sunden & Surette, 1998; Croson & Gneezy, 2009; Friesdorf, Conway & Gawronski, 2015). Yet, little is known about gender differences in the equity-efficiency trade-off.

A set of previous studies looked at the development of preferences for equity and efficiency from childhood to adolescence. The starting point of this literature is the observation that children develop preferences for equity quite early in their lives (Blake & McAuliffe, 2011; Shaw & Olson, 2012), but adults seem to prefer efficiency over equity (Charness & Rabin, 2002; Engelmann & Strobel, 2004; Capraro, Smyth, Mylona & Niblo, 2014). In agreement with this view, several authors have observed a decrease in equity preferences accompanied by an increase in efficiency considerations from 8 to 19 years old (Almås, Cappelen, Sørensen & Tungodden, 2010; Fehr, Glätze-Rützler & Sutter, 2013; Martinsson, Nordblom, Rützler & Sutter, 2011; Meuwese, Crone, de Rooij & Güroğlu, 2015). Interestingly, this increase in efficiency considerations from childhood to adolescence appears to be stronger for boys than for girls (Almås et al, 2010; Meuwese et al, 2015). Since little girls and little boys have similar preferences for equity (Blake & McAuliffe, 2011; Shaw & Olson, 2012), this differential development of preferences for efficiency from childhood to adolescence suggests that adult females might end up preferring equitable over efficient distributions to a greater extent than adult males do.

However, a series of experiments, from different research groups and using different empirical techniques, provide weak and inconclusive evidence of this prediction. Andreoni & Vesterlund (2001) conducted a series of dictator game experiments with varying cost-to-benefit ratio with 142 university students. In doing so, they found that, when the benefit of the altruistic action is greater than its cost (and thus it maximises efficiency), males give more than females; conversely, when the benefit of the altruistic action is smaller than or equal to its cost, then females give more than males. Fehr, Naef and Schmidt (2006) analysed gender differences (among university students) in situations in which the decision-maker has to choose between three allocations of money: one that maximises efficiency, one that minimises inequity, and one that maximises the payoff of the worse-off player. In doing so, they found that females are weakly significantly more egalitarian than males. These studies, however, do not explicitly pit equity against efficiency. The only study that I am aware of pitting equity against efficiency is the one of Durante, Putterman and Van der Weele (2014). They implemented a tax-game in which participants (undergraduate university students) had to choose which tax to implement in a group of twenty tax payers. They conducted several treatments and found that, overall, females are more pro-redistribution than males, even when redistribution is associated to an efficiency loss. However, in their baseline, they found no significant gender differences, but only a slight trend, according to which the average tax implemented by females was 4 percentage points higher than the average tax implemented by males.

In summary, while previous research seems to be in line with the view that adult females are more equitable than adult males, further work is needed to confirm this hypothesis. To this end, here I report the largest-to-date test of gender differences in the equity-efficiency trade-off.

I measure the equity-efficiency trade-off using the recently introduced Trade-Off Game (TOG; Capraro and Rand, 2018; Tappin & Capraro, 2018). In the TOG, one decision-maker has to choose between two allocations of money that affect the decision-maker itself as well as two other people: one allocation equalises the payoff of the three players; the other allocation maximises the sum of the payoffs of the three players. The two other people do not make any choice: they are simply paid according to the decision-maker's choice.

Here I analyse all the TOG experiments that my research group has conducted since we introduced this game. This is a large dataset containing N=5,056 observations, divided in 23 experimental treatments, collected among US based participants, recruited on Amazon Mechanical Turk (AMT).[1]

This dataset offers an excellent occasion to study gender differences in the equity-efficiency trade-off, not only because of its largeness, but also because of the variety of its experimental treatments, which allows me to explore the role of two potential moderators of theoretical and practical importance.

The first moderator is whether efficiency is aligned with self-interest. In real-life decisions, sometimes, but not always, the equitable choice is costly also for the decision-maker. Since males are known to be more self-regarding than females (Engel, 2011; Rand et al., 2016; Brañas-Garza et al., 2018), it is important to test whether gender differences in the equity-efficiency trade-off, if existing, are actually driven by gender differences in the weight that people place on self- versus other-interest. The current dataset is ideal to test for this moderator, because in 10 out of the 23 treatments (N=2,470) the equitable option is costly for the decision maker. (See the Method section for details about the exact payoffs.)

The second moderator is the frame of the trade-off game. Real-life decision problems, especially in political debate, are not formulated with a neutral language, but are often framed with a morally loaded language meant to suggest the right thing to do. Therefore, understanding whether gender differences in the equity-efficiency trade-off, if existing, depend on the moral frame of the game is of great practical interest. The current dataset is ideal also to test for this moderator, because 10 treatments (N=1,966) are framed using a language that suggest that being equitable is the right thing to do, whereas 7 treatments (N=1,769) are framed in such a way to suggest that being efficient is the right thing to do. (See the Method section for details about the frames.)

The resulting analysis demonstrates that females prefer equity over efficiency to a greater extent than males do, independently of the moral frame.

**Method**

*Measure of the equity-efficiency trade-off: The Trade-Off game*

The notions of equity and efficiency have been debated by philosophers, psychologists, and economists for centuries, leading to different positions and different definitions. Here, I follow the recent work in experimental psychology and economics, which tend to consider equity and efficiency as synonyms of equality and Pareto improvement, respectively (Almås, Cappelen,

---

[1] AMT is an online labour market that has been shown to produce reliable results on economic games (Paolacci, Chandler & Ipeirotis, 2010; Horton, Rand & Zeckhauser, 2011; Rand, 2012; Paolacci & Chandler, 2014; Arechar, Gächter & Molleman, 2018; Brañas-Garza et al., 2018). Additionally, the typical AMT sample is more heterogeneous than the classical student sample used in most laboratory experiments (Berinsky, Huber & Lenz, 2012).

Sørensen & Tungodden, 2010; Fehr, Glätze-Rützler & Sutter, 2013; Martinsson, Nordblom, Rützler & Sutter, 2011; Meuwese, Crone, de Rooij & Güroğlu, 2015). These works adopt, as a measure of the equity-efficiency trade-off, distribution games such that one decision-maker has to unilaterally choose between two allocations of money: one equalises the payoff of all the participants involved in the interaction; the other one is a Pareto improvement such that all participants are better off than in the equitable distribution, but they receive different payoffs. Along these lines, as a measure of the equity-efficiency trade-off, I adopt the Trade-Off Game (TOG; Capraro & Rand, 2018; Tappin & Capraro, 2018). In the TOG, one decision-maker has to unilaterally choose between two allocations of money that affect the decision-maker itself and two other players. One allocation equalises the payoff of the three players; the other allocation is a Pareto improvement. The other two players have no active role and they are only paid according to the decision-maker choice.

*The dataset*

I analyse N=5,056 US based participants recruited on AMT (females = 43.3%; mean age = 34.11, sd = 11.32). All these participants played the TOG. The details of the experiments depended on the particular treatment. The dataset contains 23 treatments, which can be classified in five main (non-exclusive) groups.

*Trade-Off Game in which efficiency and self-interest are aligned (10 treatments, N=2,469)*

The equitable allocation is [13 13 13], that is, each player receives $0.13; the efficient allocation is [15 23 13], that is, the decision maker receives $0.15, Player B receives $0.23, and Player C receives $0.13.

*Trade-Off Game without the previous "selfish confound" (13 treatments, N=2,605)*

The equitable allocation is [13 13 13]; the efficient allocation is [13 23 13].

*Trade-Off game with comprehension questions (9 treatments, N=1,666)*

Participants are asked two comprehension questions: (i) "What choice should you make if you want all players involved to get the same payoff?" (ii) "What choice should you make if you want to maximize the total group payoff (i.e., the sum of your bonus plus the bonuses of Players A and B)?" In the TOGs with comprehension questions, I include in the analysis only participants who responded to both comprehension questions correctly.

*Trade-Off game in the "equitable frame" (10 treatments, N=1,966)*

The equitable option is presented with a positively loaded language and/or the efficient option is presented with a negatively loaded language. The frames have been implemented in several different ways depending on the study. For example, Capraro and Rand's (2018) Study 1 labels the equitable choice as the "nice" choice and the efficient choice as the "non nice" choice. Capraro and Rand's (2018) Study 3 labels the equitable choice as "the more fair choice" and the efficient choice as the "less fair" choice. Tappin and Capraro (2018) labels the equitable option as the "fair" choice and the efficient choice as "Option 2". In the same study, another treatment labels the equitable option as "Option 1" and the efficient option as "unfair". All these manipulations had the effect of making participants more likely to choose the equitable allocation. Effect sizes were independent of the particular manipulation being used. Moreover, Capraro and Rand's (2018) Study 4 shows that this labelling technique has the effect of changing participants' perception of what is the morally right thing to do.

*Trade-Off game in the "efficient frame" (7 treatments, N=1,769)*

The efficient option is presented with a positively loaded language and/or the equitable option is presented with a negatively loaded language. The frames have been implemented in several different ways, depending on the study. For example, Capraro and Rand's (2018) Study 1 labels the efficient choice as the "nice" choice and the equitable choice as the "non nice" choice. Capraro and Rand's (2018) Study 3 labels the efficient choice as the "more generous" choice and the equitable choice as the "less generous" choice. Tappin and Capraro (2018) labels the efficient option as the "generous" choice and the equitable choice as "Option 2". In the same study, another treatment labels the efficient option as "Option 1" and the equitable option as "ungenerous". All these techniques had the effect of making participants more likely to choose the efficient allocation. Effect sizes were independent of the particular manipulation being used. Moreover, Capraro and Rand's (2018) Study 4 shows that this labelling technique has the effect of changing participants' perception of what is the morally right thing to do.

*Trade-Off game in the "neutral frame" (6 treatments, N=1,318)*

One option is called "Option 1", the other one is called "Option 2".

*Variables*

I define two individual-level variables: *equal_choice* is a dummy variable that is equal to 1 if the corresponding individual chooses the equitable option in the TOG; *female* is a self-explanatory dummy variable. Through these variables, I build the key treatment-level variables needed for the meta-analysis: for each treatment, *coeff* and *error* represent, respectively, the coefficient and the standard error of logit regression predicting *equal_choice* as a function of *female*. Apart from these two, there are three more treatment-level variables: *efficient_selfish* is a dummy variable equal to 1 if, in the corresponding treatment, the efficient choice maximises the payoff of the decision-maker; *comprehension* is a dummy variable equal to 1 if the corresponding treatment contains comprehension questions; *frame* is a categorical variable equal to 1 if the corresponding treatment is framed such that the efficient choice is presented as being the morally right thing to do, equal to -1 if the corresponding treatment is framed such that the equitable choice is presented as being the morally right thing to do, and equal to 0 if the corresponding treatment is neutrally framed.

**Results**

As a first step of the analysis, I look at the overall effect of *female* on *equal_choice*. To do so, I conduct random-effect meta-analysis using the Stata command: *metan coeff error, random*. The results, shown in Figure 1, clearly show a significant overall effect such that females prefer equity over efficiency to a greater extent than males do (effect size = 0.314, 95% CI = [0.142, 0.487], Z=3.58, p < 0.001). There is also significant evidence of heterogeneity across studies in the true size of this effect (heterogeneity chi-squared = 37.30, p = 0.022, variation in effect size attributable to heterogeneity = 41%).

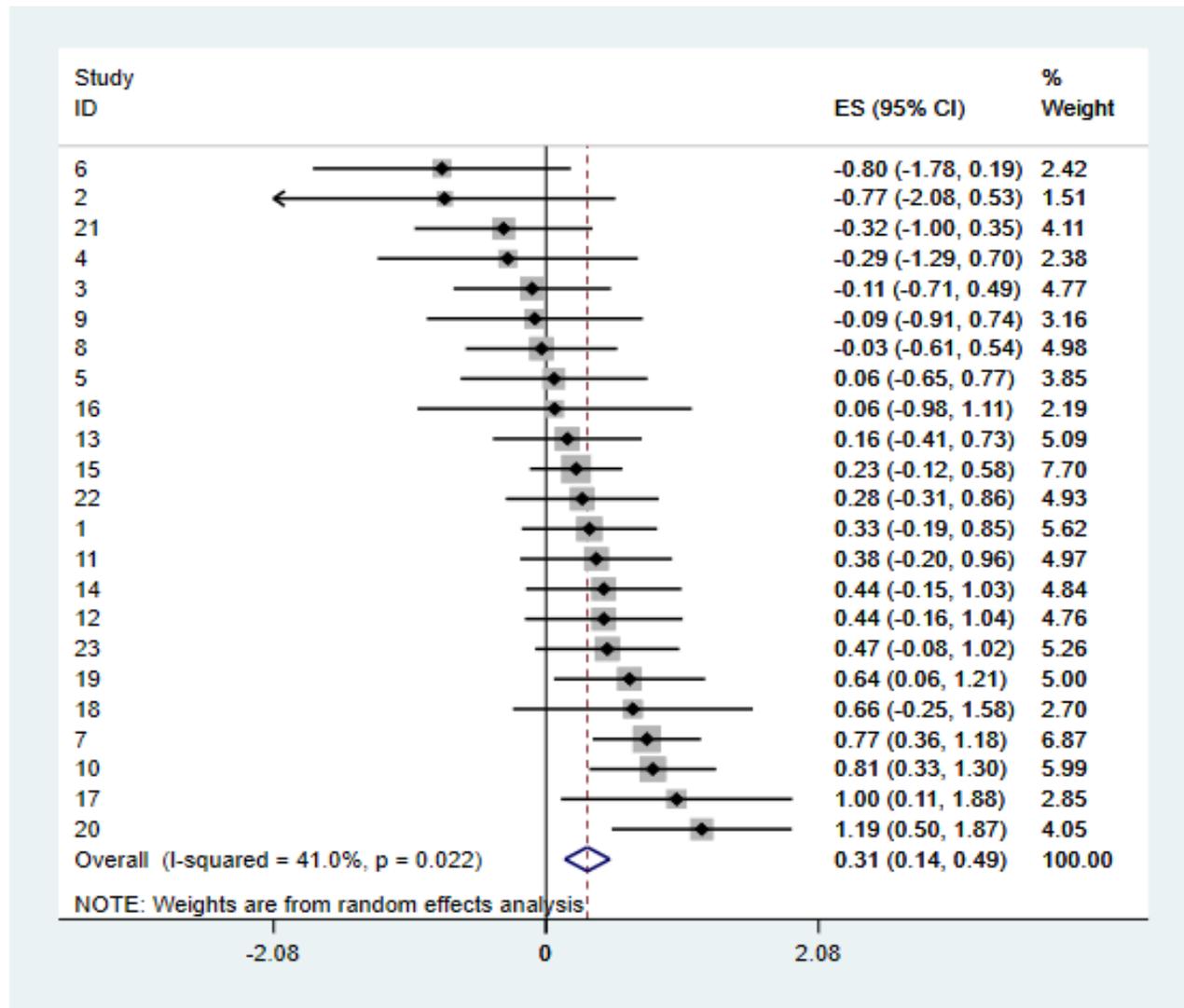

**Figure 1. Forest plot of random-effect meta-analysis of the effect of gender on Trade-Off game choice.** *Females are significantly more likely than males to choose the equitable allocation of money in the TOG.*

Next, I test whether these gender differences are robust after controlling for two potential confounds. The first possible source of confound is that, in some of the treatments, the efficient option is aligned with the payoff-maximising option. Since previous meta-analyses using the dictator game found that males are more self-interested than females (Engel, 2011; Rand et al., 2016; Brañas-Garza et al., 2018), it is in principle possible that gender differences in TOG choices are entirely driven by males being more self-regarding than females. To exclude this, I use meta-regression to test for the influence of the variable *efficient_selfish* on *coeff* – Stata command: *metareg coeff efficient_selfish, wsse(error)*. The results show that *efficient_selfish* does not significantly affect the size of the gender effect (coeff=-0.236, p=0.212). This suggests that gender differences in the TOG choices are not driven by the selfish confound. If anything, the overall effect of gender is actually numerically *lower* in the TOGs in which efficiency and self-interest are aligned (overall effect size = 0.28), compared to when they are not (overall effect size = 0.39).

The second potential source of confound is comprehension. Since some of the TOG treatments contain comprehension questions while others do not, it is crucial to make sure that gender differences in TOG choices are not driven by participants not comprehending the decision problem. To this end, I use the Stata command: *metareg coeff comprehension, wsse(error)*. The results show that comprehension does not significantly impact the size of the gender effect (p=0.285). If anything, the overall effect of gender is actually numerically *higher* in the TOGs with comprehension questions (overall effect size = 0.40), compared to those without comprehension questions (overall effect size = 0.30).

I now move to the moderating role of framing the TOG using loaded language in such a way to suggest that one option is morally better than the other one. The forest plot in Figure 2, generated through the Stata command *metan coeff error, by(frame) label(namevar=treatment)*, shows that the gender effect in TOG choices is numerically higher in the equitable frame (overall effect size = 0.46), compared to the neutral frame (overall effect size = 0.28), and the efficient frame (overall effect size = 0.22). However, the fact that the between groups heterogeneity is not significant (p=0.287) suggests that the gender effect in TOG choices does not significantly differ across TOG frames. This suggestion is confirmed by meta-regression exploring the moderating effect of *frame* on the effect size (Stata command: *metareg coeff frame, wsse(error)*; p=0.207).

Looking at gender differences in TOG choices within each frame, *metan coeff error if(frame==1)* demonstrates that, in the efficient frame, the overall gender effect is marginally significant (overall effect size = 0.224, 95% CI = [-0.027,0.475], Z=1.75, p=0.081). Similar Stata commands find that there is a significant overall gender effect on TOG choices both in the neutral frame (overall effect size = 0.283, 95% CI = [0.058,0.507]), Z=2.47, p=0.014) and in the equitable frame (overall effect size = 0.458, 95% CI = [0.264,0.652]), Z=4.64, p<0.001).

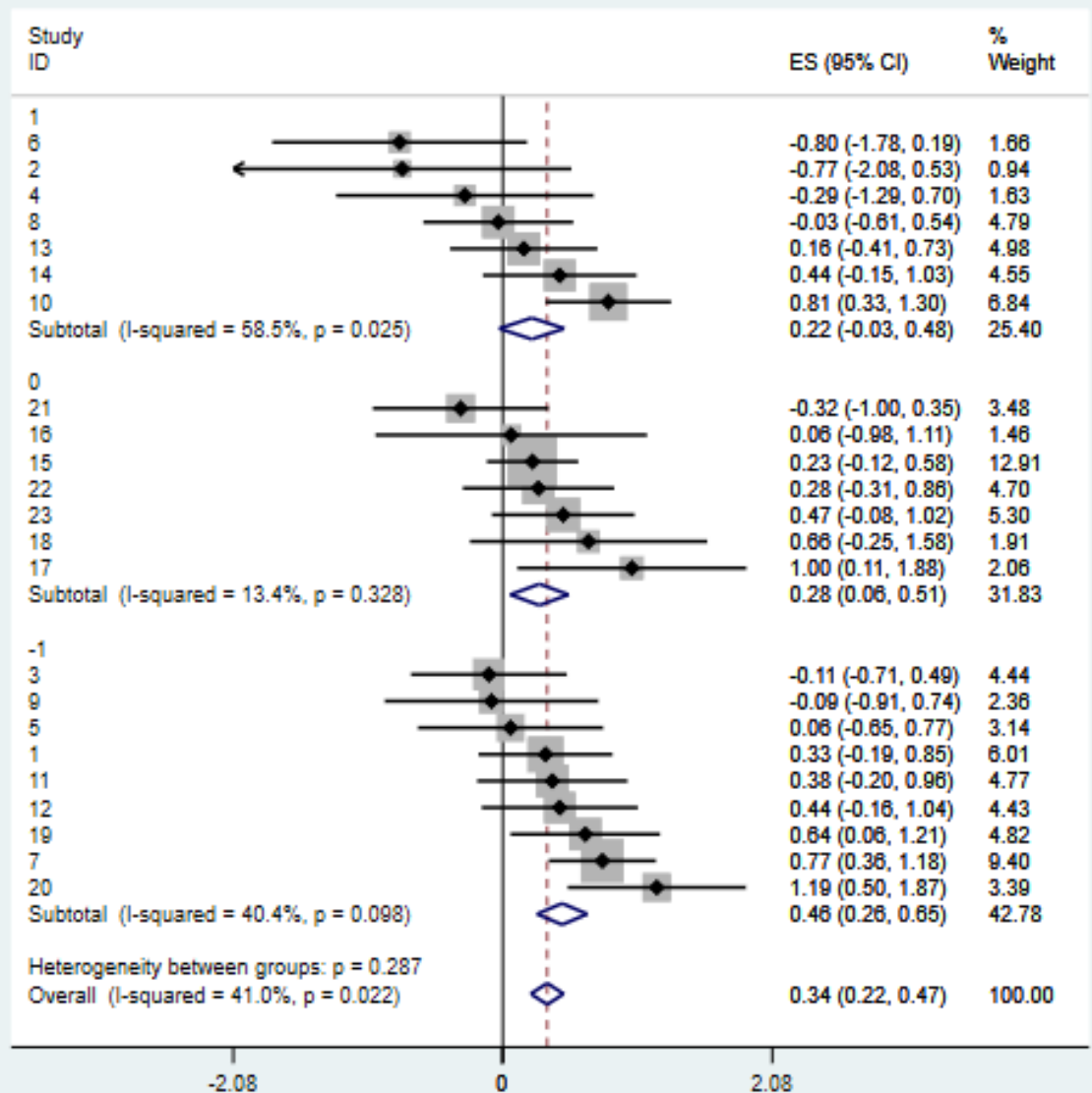

*Figure 2. Forest plot of random-effect meta-analysis of the effect of gender on Trade-Off game choice across frames. Gender differences in TOG choices do not significantly depend on the TOG frame.*

**Discussion**

To summarise, I reported the largest-to-date analysis of gender differences in the equity-efficiency trade-off: N=5,056 observations collected among US based Turkers using the Trade-Off game as a measure of the equity-efficiency trade-off (Capraro & Rand, 2018; Tappin & Capraro, 2018). The analysis provides evidence that females prefer equity over efficiency to a greater extent than males do. Importantly, the effect is not driven by situations in which efficiency is aligned with self-interest. Furthermore, the effect is relatively stable across moral frames: presenting either the equitable or the efficient option as being the morally right thing to do does not significantly changes gender differences in the equity-efficiency trade-off.

Previous work on gender differences in the equity-efficiency trade-off focused on the development of preferences for equity and efficiency from childhood to adolescence. Interestingly, two studies found that, while little girls and little boys of age < 8 tend to display similar preferences for equity, the evolution of efficiency preferences appears later in development and seem to be stronger for boys than for girls (Almås et al, 2010; Meuwese et al, 2015). This suggests that adult males might end up having different equity-efficiency trade-off than adult females. However, studies on adults provided only weak and inconclusive evidence of this prediction. Andreoni and Vesterlund (2001) found that males donate more than females when the benefit created by altruistic action is greater than its cost, and thus the altruistic action maximises efficiency. However, this study does not explicitly pit equity against efficiency, and therefore it is difficult to say whether gender differences in altruistic behaviour are actually driven by gender differences in the equity-efficiency trade-off. Fehr et al (2006) found that females are more egalitarian than males in situations in which they have to choose between three choices, one that maximises efficiency, one that minimises inequity, and one that maximises the payoff of the worse-off player. However, also this study does not explicitly pit equity against efficiency. Durante et al (2014) instead pitted equity versus efficiency and, in their baseline treatment, found only a slight non-significant trend according to which females are more equitable than males by 4 percentage points. The current work adds to this line of literature by demonstrating, in a large dataset of over five thousand observations, that females prefer equity over efficiency to a greater extent than males do. It is worth noticing that, in the neutral treatment, the effect I find corresponds to 7 percentage points, which is numerically very similar to the effect found by Durante et al (2014). Therefore, the fact that Durante et al (2014) found only a non-significant trend was probably due to their small sample size, which limited their ability to detect relatively small effect sizes. In fact, another strength of the current work is the large sample size, which consents to detect small effect sizes and to minimise the presence of false positives. This is also a valuable point, especially in light of the current Replicability Crisis (Open Science Collaboration, 2015; Camerer et al., 2018).

Related to the current study is also the literature on gender differences in moral judgments in hypothetical moral dilemmas. Previous research suggests that females are more averse than males to physically harm one person for the greater good (Fumagalli et al, 2010; Friesdorf et al, 2015; Capraro & Sippel, 2017). Although related, the current study differs from this line of research in two main dimensions: first, it involves no physical harm (and no economic harm); second, it regards actual behaviour and not moral judgments in hypothetical dilemmas.

This work has, nevertheless, several limitations. The first one is that the dataset does not contain observations in which equity benefits the decision-maker. Since males are known to be more self-regarding than females (Engel, 2011; Rand et al., 2016; Brañas-Garza et al., 2018), the obvious prediction is that gender differences in the equity-efficiency trade-off would decrease, and perhaps even reverse, if equity becomes beneficial for the decision-maker. Future work could test this hypothesis. Similarly, a second limitation is that the dataset does not contain observations in which efficiency harms one of the players involved in the interaction. Exploring gender differences in this situation would be an important extension, because, in reality, efficiency often harms worse-off players. Previous research suggests that females are more averse than males to physically harm one person for the greater good (Fumagalli et al, 2010; Friesdorf et al, 2015; Capraro & Sippel, 2017). Although it is not obvious that gender differences in physical harm map onto gender differences in economic harm, this might suggest that, in case efficiency harms one of the players, gender differences in the equity-efficiency trade-off might increase. Future research could test this prediction. The third limitation regards the stakes of the TOG, which, in all the treatments, are relatively small. Previous work suggests that stakes have no effect on people's behaviour in a number of economic games involving pro-sociality, at least when stakes are not too high (Forsythe et al., 1994; Carpenter, Verhoogen, & Burks, 2005; Johansson-Stenman, Mahmud, & Martinsson, 2005; Brañas-Garza et al., 2018; Larney, Rotella, & Barclay, 2019); other studies have indeed found evidence that pro-sociality decreases at very high stakes (Carpenter et al., 2005; Andersen, Ertaç, Gneezy, Hoffman, & List, 2011). To the best of my knowledge, there have been no studies exploring the stake effect on the equity-efficiency trade-off. Testing whether this trade-off varies as a function of the stakes and, if it does so, testing how this variation interacts with gender is an interesting direction for future research. The fourth limitation is that this dataset does not allow to answer the question of why females prefer equity over efficiency to a greater extent than males do. At this stage of research, I can only speculate. An influential line of literature suggests that gender differences in behaviour are partly due to the different roles that males and females tend to occupy in society (Eagly, 1987; Eagly & Wood, 1999). Along these lines, one potential explanation for the current findings is that, from childhood to adolescent, females and males start differentiating their social roles, with females going to cover, on average, roles involving resource distributions, while males going to cover, on average, roles involving the creation of resources. With such a role division, it would be optimal to create resources efficiently (in order to maximise the total resource to be distributed) and then divide them equitably (in order to minimise within-group conflicts). This logic could explain why adult males and females tend to display, on average, different preferences for resource distribution. Exploring this and potentially other explanations is an important avenue for future research.

In sum, this work shows that females, on average, prefer equity over efficiency to a greater extent than males do. Future work should explore the causes and the boundaries of this effect.